\documentclass{emulateapj} 
\usepackage{natbib} 
\usepackage{url} 
\usepackage[usenames,dvipsnames]{color} 
\usepackage{graphicx} 
\usepackage{multirow}
\usepackage{enumitem}
\usepackage{textcomp}
\usepackage{wasysym}
\usepackage{hyperref}
\setlist{noitemsep} 
\slugcomment{version 0.0} 
\shortauthors{A.~Bilous et al.}

\newcommand{\src}{SAX J1810.8$-$2609}

\begin{document}

\title{Millisecond oscillations in the bursting flux of \src}
 
\author{A.~V.~Bilous\altaffilmark{1}}
\author{A.~L.~Watts\altaffilmark{1}}
\author{D.~K.~Galloway\altaffilmark{2,3}}
\author{J.~J.~M.~in ’t Zand\altaffilmark{4}}

\altaffiltext{1}{Anton Pannekoek Institute for Astronomy, 
                 University of Amsterdam, Science Park 904, 
                 1098 XH Amsterdam, The Netherlands \email{A.Bilous@uva.nl}}
\altaffiltext{2}{School of Physics \& Astronomy, Monash University, Clayton, VIC 3800, Australia}
\altaffiltext{3}{Monash Centre for Astrophysics, Monash University, Clayton, VIC 3800, Australia}
\altaffiltext{4}{SRON Netherlands Institute for Space Research, Sorbonnelaan 2, 3584 CA Utrecht, the Netherlands}
                 
\begin{abstract}
\src\ is a faint X-ray transient, mostly known for its abnormally low quiescent thermal luminosity, which disagrees with  
standard cooling models. It is also one of a small sample of stars whose mass and radius have been estimated using spectral 
modeling of one of its thermonuclear bursts. 
Here we report the discovery of millisecond oscillations in a type I thermonuclear X-ray burst from \src\ 
observed by \textit{RXTE} during the 2007 outburst. A strong signal (Leahy-normalized power of 71.5, $4.5\times10^{-9}$ chance of 
coincidence with a conservative estimate for the number of trials) was present at 531.8\,Hz during the decay of one
out of six bursts observed. Oscillations were detected for about 6 seconds, during which their frequency 
increased from 531.4 to 531.9\,Hz in a manner similar to other burst oscillation sources. 
The millisecond oscillations discovered pinpoint the spin frequency of the neutron star, which is important for
the spectral modeling, associated mass-radius inference, and understanding the evolutionary status and cooling behavior of the star.
As of April 2018 the source is in outburst again, providing a fleeting opportunity to acquire
new material for the burst oscillation searches.
\end{abstract}
   

\maketitle

\section{Introduction}

Millisecond oscillations in type I X-ray bursts are caused by the development of asymmetric bright patches during thermonuclear 
explosions on the surface of accreting neutron stars. Burst oscillations are rare phenomena, having only been observed in 18 
out of about 110 sources\footnote{\url{http://www.sron.nl/~jeanz/bursterlist.html}, see also \citet{Galloway2008}, \citet{Watts2012}, 
and references therein.}.  Oscillation amplitudes can take a range of values (typically, 5\%--20\% for the fractional rms amplitudes),
with higher amplitudes being more prevalent in certain accretion states and the majority of the non-detections can 
be attributed to the lack of observations in the appropriate state \citep{Ootes2017}.  However what causes this spread of amplitudes
remains a mystery.

Burst oscillations can occur in any part of the burst and usually last for several seconds. They are highly coherent, 
with frequencies typically (but not always)
drifting smoothly upwards by 1--3\,Hz towards the asymptotic maximum, nearly constant for each source. Oscillation 
frequencies range from 245 to 620\,Hz
\citep{Watts2012}, but there have also been detected oscillations at frequencies as low as 11\,Hz \citep[e.g.][]{Cavecchi2011}. 
The similarity between the oscillation and spin frequencies of several accretion-powered pulsars 
(e.g. SAX J1808.4$-$3658 \citep{Chakrabarty2003}, XTE J1814$-$338 \citep{Strohmayer2003}, and other) revealed the tight connection between 
the oscillation frequency and the spin frequency of the neutron star, enabling the determination of spin frequencies for several 
neutron stars which do not manifest themselves
as pulsars.

\begin{table*}
\begin{center} 
\caption{Summary of burst properties. All quantities except for burst duration are taken from the MINBAR catalog. Estimates of 
burst duration come from Bilous et al. (in prep). \label{table:data}}
\begin{tabular}{cccccccccc} 
\hline\\ 
\parbox{0.4cm}{\centering Burst \# } &
\parbox{1.2cm}{\centering \textit{RXTE} ObsID} &
\parbox{1.5cm}{\centering  Burst epoch (MJD UT)} &
\parbox{1.9cm}{\centering  Burst start date\\(yyyy-mm-dd)} &
\parbox{1.2cm}{\centering  Burst duration (s)} &
\parbox{1.8cm}{\centering  Peak count rate per PCU ($10^3$\,ct\,s$^{-1}$)} &
\parbox{1.9cm}{\centering  Bolometric fluence\\($10^{-6}$\,erg\,cm$^{-2}$)} &
\parbox{2.2cm}{\centering  Persistent flux\\($10^{-9}$\,erg\,cm$^{-2}$)} &
\parbox{1.5cm}{\centering  Photospheric radius expansion?} 
\\ [0.01cm]
\hline\\
1 & 93044-02-04-00 & 54325.89373 & 2007-08-13 & 106 & 6.7(1) & 1.37(1) & 1.012(5) & yes \\
2 & 93044-02-05-00 & 54326.97236 & 2007-08-14 & 89 & 5.6(1) & 1.05(5) & 1.225(5) & no \\
3 & 93044-02-07-00 & 54332.87613 & 2007-08-20 & 75 & 8.8(1) & 0.81(1) & 1.12(6) & yes \\
4 & 93093-01-01-00 & 54369.80255  & 2007-09-26 & 91 & 5.8(1) & 0.96(1) & 1.21(3) & no \\
5 & 93093-01-01-00 & 54370.05022 & 2007-09-27 & 70 & 5.5(1) & 0.92(1) & 1.21(3) & no \\
6 & 93093-01-01-01 & 54370.33383 & 2007-09-27 & 66 & 5.8(1) & 0.93(9) & 1.16(8) & no\\
\hline 
\end{tabular} 
\end{center}
\end{table*}

Burst oscillations are unique tools for exploring the nuclear burning in the strong gravity/magnetic fields on the neutron star surface. 
Also, folded waveforms of burst oscillations bear the imprints of the gravitational 
field at the neutron star surface, allowing for simultaneous measurements of the star's mass and radius and 
thus constraining the equation of state of matter at supra-nuclear densities \citep[see][for an overview]{Watts2016}.

\begin{figure*}
 \centering
 \includegraphics[scale=0.975]{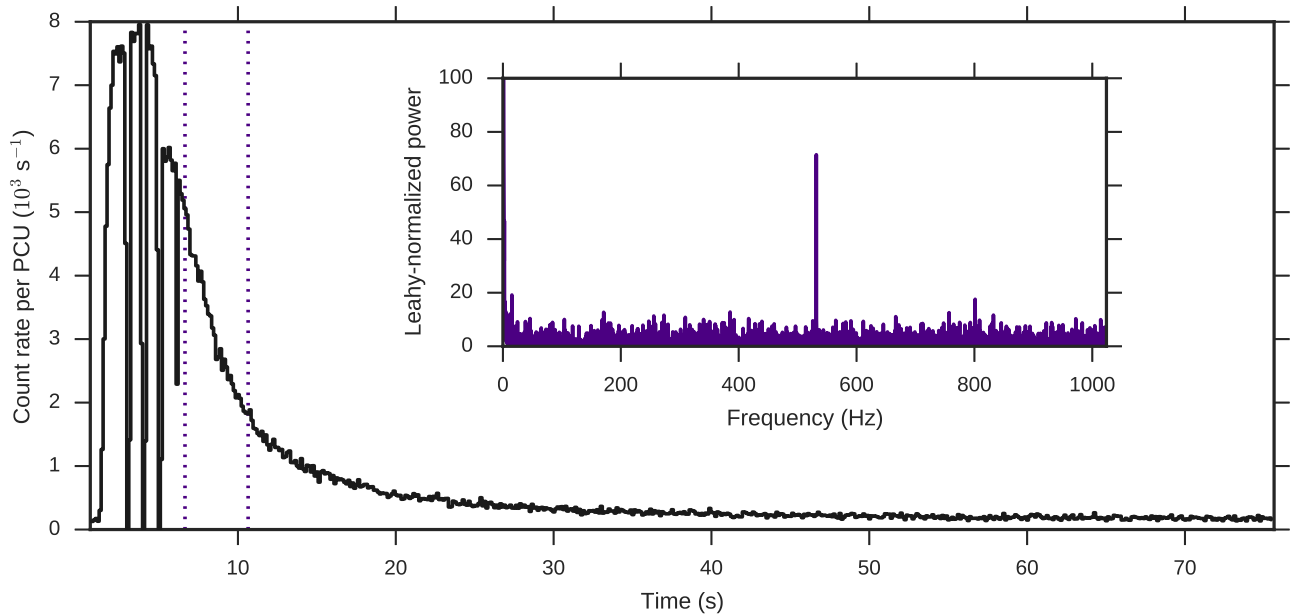}
 \caption{Burst \#3 from Table~\ref{table:data}. The main panel shows count rate of Scientific Event data, with time 
 counted from the burst start. Small data gaps due to telemetry limitations are present at higher count rates. The inset
 shows the Leahy-normalized power spectrum from the 4-s time interval marked with dashed lines on the main panel.
 In this time interval the oscillation signal was the strongest, with the power of over 70 at 531.75\,Hz.}
 \label{fig:FFT}
\end{figure*}

\src\ is a low-mass X-ray binary (LMXB) discovered with the Wide Field Camera on board the BeppoSAX satellite \citep{Ubertini1998}. 
The source
spends most of time in a quiescent state, with three outbursts observed so far: in 1998, 2007 and 
2012. As of April 2018, the source is reported to be in outburst again \citet{Negoro2018}.
Both quiescent and outburst luminosities of \src\ are on the lower end of the corresponding luminosity distributions for 
neutron star X-ray transients\footnote{However, \citet{Degenaar2013} argued that the flux of the 2012 outburst
suggests that \src\ may be a bright transient fortuitously observed previously during its fainter outbursts.}.
Its thermal luminosity during quiescence is in disagreement with NS heating and cooling models, suggesting 
an order of magnitude smaller accretion rate than is inferred from its outburst activity
\citep{Jonker2004,Allen2018}. 
Thermal emission from one of the Type I bursts from the 2007 outburst has been used by \citet{Suleymanov2017}
to constrain mass and radius of this neutron star using the direct cooling tail method,
yielding $R = 11.5$--$13.0$\,km for the 99\% confidence region at an assumed mass of $M = 1.3$--$1.8$\,$M_\odot$.
However the 68\% confidence upper limit that they find on the mass is $\sim 1.5$\,$M_\odot$.

Knowing the spin frequency of NS is important for spectral modeling both in quiescence and during 
outbursts and bursts \citep{Burke2018,Baubock2015}. It would also put constraints on the evolutionary history of the system,
which could be helpful for explaining the low temperature of the NS \citep{Allen2018}. 

In this letter we report the discovery of 531.8\,Hz burst oscillations in one of the six bursts recorded by Rossi X-Ray 
Timing Explorer (\textit{RXTE}) in the 2007 outburst.
This makes \src\ the 19th known source with burst oscillations and places the system among the fast-spinning NS LMXBs.

\section{Data analysis}
We analyzed the data from the Proportional Counter Array on board of \textit{RXTE} for the six bursts which were
observed between August and September 2007. The observation IDs and the MJDs of arrival (Table~\ref{table:data})
were taken from the Multi-INstrument Burst ARchive 
(MINBAR\footnote{\url{https://burst.sci.monash.edu/minbar/}}, Galloway et al., in prep).
The duration was defined as the time span where 0.5-s count rate from Standard-2 files was larger than mean plus two standard deviations 
of the count rate in the pre-burst baseline window.
The Science Event mode was used, with time resolution of 122\,$\mu$s and no energy cuts.

In order to search for oscillations, we applied the standard technique of calculating power spectra in sliding 
windows of $\Delta T=0.5$, 1, 2 and 4 seconds, each new window starting with a 0.5-s offset with respect to the previous one. 
For each window, Fourier frequencies between 2 and 2002\,Hz were recorded. The upper limit on the oscillation frequency reflects the upper limit on NS spin 
frequency set by all current reasonable models of the neutron star equation of state \citep{Haensel2009}.

One of the bursts (\#3 in Table~\ref{table:data}) yielded a strong 
signal at frequencies of about 532\,Hz for all FFT window lengths. The Leahy-normalized power \citep{Leahy1983} was largest
for 4-s windows, reaching $P_\mathrm{m}=71.5$ at 531.75\,Hz. Assuming the noise power is distributed as $\chi^2$ with two degrees of freedom,
the single-trial probability of obtaining such power is $3\times10^{-16}$. By a conservative estimate, counting all time bins and time 
windows as independent trials, the number of trials
for all six bursts in total include $994\times4$ time windows with 1000--8000 frequency bins per window, summing up to 
$N_\mathrm{tr}=994\times(1000+2000+4000+8000)\approx1.5\times 10^7$.
Even with this conservative estimate the chance probability of obtaining such a strong signal ($4.5\times 10^{-9}$)  is negligible.
The estimated value of the chance probability corresponds to $5.75\sigma$ of the normal distribution.

Up to now, there have been known 18 sources with burst oscillations. Nine more sources have tentative detections \citep{Watts2012}. Normally,
detection of coherent oscillations at similar frequencies in multiple bursts or in multiple independent 
time bins serves as a firm corroboration of burst oscillations. The oscillations from \src\ were detected in one burst only,
motivating more searches for burst oscillations in the future, perhaps during the current outburst.  

For burst \#3, oscillations with $P_\mathrm{m}>24$ (corresponding to $p<6\times 10^{-6}$ 
for $\chi^2$ noise distribution) were detected in two independent consecutive 4-s time bins. The probability of such a detection being due to chance
is $p^2\times 994\times 8000\approx 3 \times 10^{-4}$,  
where we make the most conservative estimate for the number of trials, $N_\mathrm{tr} = 994\times 2000\times \Delta T$. 
Overall, the $P_\mathrm{m}=71.5$, $5.75\sigma$ single-bin single burst oscillation detection 
for \src\ is more significant than detections from other sources deemed ``tentative'' in \citet{Watts2012} (up to $4.9\sigma$).
It is also more significant than at least some of the discovery detections of subsequently confirmed burst oscillations 
\citep[e.g. SAX J1750.8$-$2900, $5.0\sigma$, ][]{Kaaret2002}.

\begin{figure}
 \centering
 \includegraphics[scale=0.975]{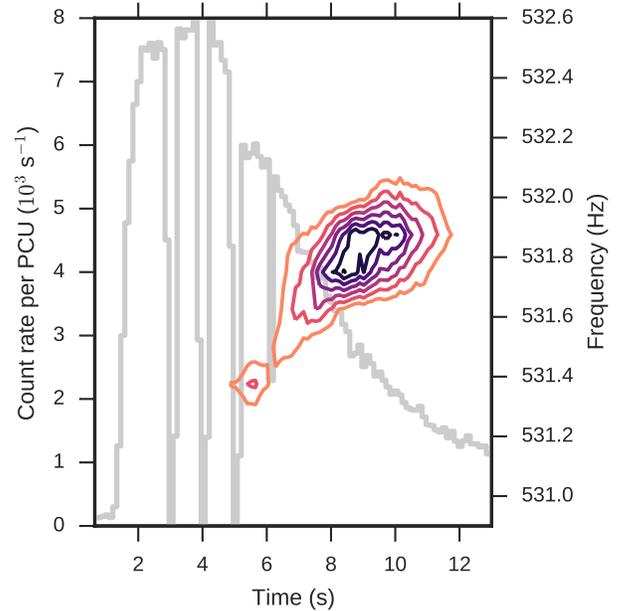}
 \caption{$Z^2$ power spectrum of burst \#3. The $Z^2$ values were computed in 4-s intervals overlapping by 3.875 seconds,
at frequencies oversampled by a factor of 2. The power is plotted at the midpoint of each interval.
Contour levels mark $Z^2$ from 20 to 80 with the step of 10. The peak power was 81.}
 \label{fig:Z2}
\end{figure}

In order to explore the possible frequency drift, we computed a dynamic power spectrum using $Z^2$ statistics \citep{Buccheri1983}.
Unlike Fourier transforms which use binned data, $Z^2$ statistics use the time of arrival of each individual photon 
and can be computed at arbitrarily close frequencies (although the frequency resolution is still determined by the choice of $\Delta T$).
We used 4-s time bins overlapping by 3.875\,s and frequency bins starting from 531\,Hz and increasing in 0.125\,Hz steps. The dynamic
power spectrum is shown on Fig.~\ref{fig:Z2}. The oscillation signal is present at 6--12\,s counting from the burst start
and the frequency drifts from 531.4 to about 531.9\,Hz. The largest value of the $Z^2$ statistic was 81.

The Leahy-normalized power spectra were used to compute the fractional amplitude of oscillations \citep{Watts2005}:
\begin{equation}
\label{eq:framp}
 A = \left(\frac{P_\mathrm{s}}{ N_\mathrm{m} }\right)^{1/2}\frac{N_\mathrm{m}}{N_\mathrm{m}-N_\mathrm{bkg}}. 
\end{equation}
Here $P_\mathrm{s}$ is the Leahy-normalized power of signal in the absence of noise, $N_\mathrm{m}$ is the number of 
photons in the given time bin and $N_\mathrm{bkg}$ is the estimated number of background photons in the same time bin. 
We used the median value of $P_\mathrm{s} = P_\mathrm{m}+1$ 
from the distribution of $P_\mathrm{s}$ given $P_\mathrm{m}$ 
derived by \cite{Groth1975}, but with Leahy normalization \citep[see discussion in][]{Watts2012}. 
The uncertainty on $P_\mathrm{s}$ was taken from [0.159, 0.841] percentiles of the same distribution.
The uncertainty on the number of 
photons in a time bin, $N_\mathrm{m}$, was taken to be Poissonian and 
the uncertainty in the background level was taken to be the standard deviation of count rates in the 4-s overlapping time bins within 120-s
window prior to the burst onset. Fractional amplitude errors were calculated as linear error propagation of the independent parameters
\citep{Ootes2017}. For the strongest signal, the fractional rms amplitude was $4.7\pm 0.6$\%. We did not detect any signal at the 
first harmonic frequency: between 1063 and 1064\,Hz the maximum $P_\mathrm{m}$ was 2.6.

No signal was found in any of the other five bursts, although their peak count rates are comparable (Table~\ref{table:data}).
Bolometric fluences and the levels of persistent flux at burst times are also similar between all six bursts, 
although burst \#3 has the largest peak count rate and the  smallest bolometric fluence. 
Interestingly though, burst \#3 is one of the two bursts out of six with photospheric radius expansion in the MINBAR catalog.  

\section{Discussion}

The oscillations from \src\ have properties typical for the burst oscillations from other sources: frequency around 500\,Hz, duration of few seconds,
the small upward frequency drift towards an asymptotic frequency and moderate fractional amplitudes. They have been observed in the burst
with strongest  photospheric radius expansion (out of two such bursts in the sample).

Another burst with PRE, \#1 from Table~\ref{table:data} has been used to constrain mass and radius of this neutron star using the 
direct cooling tail method, 
which uses atmospheric models to convert the spectral evolution during burst tail to the stellar angular size
\citep{Suleymanov2017}. 
As shown by \citet{Baubock2015}, rapid rotation can have a significant effect on the radius inferred using this method: 
failure to include rotational effects leads to the radius or mass being under-estimated. Neither \citet{Natilla2016} nor \citet{Suleymanov2017}
included rotation in their models, since it complicates the computations 
by introducing two more free parameters (spin period and inclination).   Establishing the spin 
frequency of \src\ allows us to make corrections to the model and obtain better constraints on mass and radius of this NS.  
A spin rate of 532 Hz could result in a radius up to $\sim 5$\% larger \citep{Baubock2015}. 

Knowing spin of NS is important for modeling the spectrum of persistent outburst emission.
\citet{Burke2018} analyzed a sample of sources where the spin is known and 
found that Comptonization strength is larger for more rapidly spinning stars.
The observations are thus in agreement with the theoretical scenario, in which for more rapidly spinning neutron stars
less energy is liberated during the deceleration of accreted material in a boundary layer,
resulting in a lower seed photon luminosity and less Compton cooling in the corona. \src\ has a relatively
large spin frequency which might naturally explain the very hard nature of its spectrum \citep{Natalucci2000}.

Finally, the high spin frequency of \src\ may corroborate or eliminate some mechanisms suggested to explain its 
very low quiescent luminosity. According to \citet{Allen2018}, some of the possible explanations
for the unusually low temperature of this NS include some enhanced cooling processes (e.g. direct Urca), 
a hybrid crust, or overestimation of the time-averaged outburst accretion rate. 
If the system is young or had extremely low accretion rates,
it could not accrete enough material to replace the NS crust during its lifetime, forming so-called hybrid crust.
Deep crustal heating is suppressed in a hybrid crust \citep{Wijnands2013}. 
The high spin frequency of \src, if due to accretion-induced spin-up, may
be at odds with the system being either young
or having low accretion rate. 

Enhanced cooling via direct Urca process requires more massive NS (1.6--1.8$M_\odot$).
Currently this mass is outside the 68\% probability region of \citet{Suleymanov2017}, however it needs to be revisited with 
spin corrections included. As was shown by \citet{Baubock2015}, for a known radius, neglecting rotation underestimates the mass.

\section*{Acknowledgements}

AVB and ALW acknowledge support from ERC Starting Grant No. 639217 CSINEUTRONSTAR (PI Watts). 
The MINBAR project acknowledges the support of the Australian Academy of Science’s Scientific Visits to Europe program, 
and the Australian Research Council’s Discovery Projects (project DP0880369) and Future Fellowship (project FT0991598) 
schemes. The research leading to these results has received funding from the European Union’s Horizon 2020 Programme 
under AHEAD project (grant agreement No. 654215).

\bibliographystyle{apj} 
\bibliography{SAX_J1810_lit}

\begin{thebibliography}{}
\expandafter\ifx\csname natexlab\endcsname\relax\def\natexlab#1{#1}\fi

\bibitem[{{Allen} {et~al.}(2018){Allen}, {Homan}, {Chakrabarty}, \&
  {Nowak}}]{Allen2018}
{Allen}, J.~L., {Homan}, J., {Chakrabarty}, D., \& {Nowak}, M. 2018, \apj, 854,
  58

\bibitem[{{Baub{\"o}ck} {et~al.}(2015){Baub{\"o}ck}, {{\"O}zel}, {Psaltis}, \&
  {Morsink}}]{Baubock2015}
{Baub{\"o}ck}, M., {{\"O}zel}, F., {Psaltis}, D., \& {Morsink}, S.~M. 2015,
  \apj, 799, 22

\bibitem[{{Buccheri} {et~al.}(1983){Buccheri}, {Bennett}, {Bignami}, {Bloemen},
  {Boriakoff}, {Caraveo}, {Hermsen}, {Kanbach}, {Manchester}, {Masnou},
  {Mayer-Hasselwander}, {{\"O}zel}, {Paul}, {Sacco}, {Scarsi}, \&
  {Strong}}]{Buccheri1983}
{Buccheri}, R., {Bennett}, K., {Bignami}, G.~F., {et~al.} 1983, \aap, 128, 245

\bibitem[{{Burke} {et~al.}(2018){Burke}, {Gilfanov}, \& {Sunyaev}}]{Burke2018}
{Burke}, M.~J., {Gilfanov}, M., \& {Sunyaev}, R. 2018, \mnras, 474, 760

\bibitem[{{Cavecchi} {et~al.}(2011){Cavecchi}, {Patruno}, {Haskell}, {Watts},
  {Levin}, {Linares}, {Altamirano}, {Wijnands}, \& {van der
  Klis}}]{Cavecchi2011}
{Cavecchi}, Y., {Patruno}, A., {Haskell}, B., {et~al.} 2011, \apjl, 740, L8

\bibitem[{{Chakrabarty} {et~al.}(2003){Chakrabarty}, {Morgan}, {Muno},
  {Galloway}, {Wijnands}, {van der Klis}, \& {Markwardt}}]{Chakrabarty2003}
{Chakrabarty}, D., {Morgan}, E.~H., {Muno}, M.~P., {et~al.} 2003, \nat, 424, 42

\bibitem[{{Degenaar} \& {Wijnands}(2013)}]{Degenaar2013}
{Degenaar}, N., \& {Wijnands}, R. 2013, in IAU Symposium, Vol. 291, Neutron
  Stars and Pulsars: Challenges and Opportunities after 80 years, ed. J.~{van
  Leeuwen}, 141--144

\bibitem[{{Galloway} {et~al.}(2008){Galloway}, {Muno}, {Hartman}, {Psaltis}, \&
  {Chakrabarty}}]{Galloway2008}
{Galloway}, D.~K., {Muno}, M.~P., {Hartman}, J.~M., {Psaltis}, D., \&
  {Chakrabarty}, D. 2008, \apjs, 179, 360

\bibitem[{{Groth}(1975)}]{Groth1975}
{Groth}, E.~J. 1975, \apjs, 29, 285

\bibitem[{{Haensel} {et~al.}(2009){Haensel}, {Zdunik}, {Bejger}, \&
  {Lattimer}}]{Haensel2009}
{Haensel}, P., {Zdunik}, J.~L., {Bejger}, M., \& {Lattimer}, J.~M. 2009, \aap,
  502, 605

\bibitem[{{Jonker} {et~al.}(2004){Jonker}, {Wijnands}, \& {van der
  Klis}}]{Jonker2004}
{Jonker}, P.~G., {Wijnands}, R., \& {van der Klis}, M. 2004, \mnras, 349, 94

\bibitem[{{Kaaret} {et~al.}(2002){Kaaret}, {in 't Zand}, {Heise}, \&
  {Tomsick}}]{Kaaret2002}
{Kaaret}, P., {in 't Zand}, J.~J.~M., {Heise}, J., \& {Tomsick}, J.~A. 2002,
  \apj, 575, 1018

\bibitem[{{Leahy} {et~al.}(1983){Leahy}, {Elsner}, \& {Weisskopf}}]{Leahy1983}
{Leahy}, D.~A., {Elsner}, R.~F., \& {Weisskopf}, M.~C. 1983, \apj, 272, 256

\bibitem[{{Natalucci} {et~al.}(2000){Natalucci}, {Bazzano}, {Cocchi},
  {Ubertini}, {Heise}, {Kuulkers}, {in 't Zand}, \& {Smith}}]{Natalucci2000}
{Natalucci}, L., {Bazzano}, A., {Cocchi}, M., {et~al.} 2000, \apj, 536, 891

\bibitem[{{N{\"a}ttil{\"a}} {et~al.}(2016){N{\"a}ttil{\"a}}, {Steiner},
  {Kajava}, {Suleimanov}, \& {Poutanen}}]{Natilla2016}
{N{\"a}ttil{\"a}}, J., {Steiner}, A.~W., {Kajava}, J.~J.~E., {Suleimanov},
  V.~F., \& {Poutanen}, J. 2016, \aap, 591, A25

\bibitem[{{Negoro} {et~al.}(2018){Negoro}, {Mihara}, {Nakahira}, {Yatabe},
  {Takao}, {Matsuoka}, {Kawai}, {Sugizaki}, {Tachibana}, {Morita}, {Sakamoto},
  {Serino}, {Sugita}, {Kawakubo}, {Hashimoto}, {Yoshida}, {Nakajima},
  {Sakamaki}, {Maruyama}, {Ueno}, {Tomida}, {Ishikawa}, {Sugawara}, {Isobe},
  {Shimomukai}, {Ueda}, {Tanimoto}, {Morita}, {Yamada}, {Tsuboi}, {Iwakiri},
  {Sasaki}, {Kawai}, {Sato}, {Tsunemi}, {Yoneyama}, {Yamauchi}, {Hidaka},
  {Iwahori}, {Kawamuro}, {Yamaoka}, \& {Shidatsu}}]{Negoro2018}
{Negoro}, H., {Mihara}, T., {Nakahira}, S., {et~al.} 2018, The Astronomer's
  Telegram, 11593

\bibitem[{{Ootes} {et~al.}(2017){Ootes}, {Watts}, {Galloway}, \&
  {Wijnands}}]{Ootes2017}
{Ootes}, L.~S., {Watts}, A.~L., {Galloway}, D.~K., \& {Wijnands}, R. 2017,
  \apj, 834, doi:10.3847/1538-4357/834/1/21

\bibitem[{{Strohmayer} {et~al.}(2003){Strohmayer}, {Markwardt}, {Swank}, \&
  {in't Zand}}]{Strohmayer2003}
{Strohmayer}, T.~E., {Markwardt}, C.~B., {Swank}, J.~H., \& {in't Zand}, J.
  2003, \apjl, 596, L67

\bibitem[{{Suleimanov} {et~al.}(2017){Suleimanov}, {Poutanen},
  {N{\"a}ttil{\"a}}, {Kajava}, {Revnivtsev}, \& {Werner}}]{Suleymanov2017}
{Suleimanov}, V.~F., {Poutanen}, J., {N{\"a}ttil{\"a}}, J., {et~al.} 2017,
  \mnras, 466, 906

\bibitem[{{Ubertini} {et~al.}(1998){Ubertini}, {in 't Zand}, {Tesseri},
  {Ricci}, \& {Piro}}]{Ubertini1998}
{Ubertini}, P., {in 't Zand}, J., {Tesseri}, A., {Ricci}, D., \& {Piro}, L.
  1998, \iaucirc, 6838

\bibitem[{{Watts}(2012)}]{Watts2012}
{Watts}, A.~L. 2012, \araa, 50, 609

\bibitem[{{Watts} {et~al.}(2005){Watts}, {Strohmayer}, \&
  {Markwardt}}]{Watts2005}
{Watts}, A.~L., {Strohmayer}, T.~E., \& {Markwardt}, C.~B. 2005, \apj, 634, 547

\bibitem[{{Watts} {et~al.}(2016){Watts}, {Andersson}, {Chakrabarty}, {Feroci},
  {Hebeler}, {Israel}, {Lamb}, {Miller}, {Morsink}, {{\"O}zel}, {Patruno},
  {Poutanen}, {Psaltis}, {Schwenk}, {Steiner}, {Stella}, {Tolos}, \& {van der
  Klis}}]{Watts2016}
{Watts}, A.~L., {Andersson}, N., {Chakrabarty}, D., {et~al.} 2016, Reviews of
  Modern Physics, 88, 021001

\bibitem[{{Wijnands} {et~al.}(2013){Wijnands}, {Degenaar}, \&
  {Page}}]{Wijnands2013}
{Wijnands}, R., {Degenaar}, N., \& {Page}, D. 2013, \mnras, 432, 2366

\end{thebibliography}


\end{document}